\documentclass[conference]{IEEEtran}
\IEEEoverridecommandlockouts
\usepackage{cite}
\usepackage{amsmath,amssymb,amsfonts}
\usepackage{algorithmic}
\usepackage{graphicx}
\usepackage{textcomp}
\usepackage{xcolor}
\usepackage{makecell}
\def\BibTeX{{\rm B\kern-.05em{\sc i\kern-.025em b}\kern-.08em
    T\kern-.1667em\lower.7ex\hbox{E}\kern-.125emX}}
    
\usepackage{enumerate}
\usepackage{tikz}
\usetikzlibrary{arrows,shapes.gates.logic.US,shapes.gates.logic.IEC,calc}
\usepackage{hyperref}
\usepackage{aurical}
\usepackage[T1]{fontenc}
\usepackage{xcolor}

\usepackage{amsmath}
\usepackage{subcaption}
\usepackage{graphicx}
\usepackage{color,soul}
\usepackage{url}
\usepackage{hyperref}
\begin{document}

\title{Actor-Critic Learning Based QoS-Aware Scheduler for Reconfigurable Wireless Networks}

\author{\IEEEauthorblockN{ Shahram Mollahasani}
\IEEEauthorblockA{\textit{School of Electrical Engineering and Computer Science} \\
\textit{University of Ottawa}\\
smollah2@uottawa.ca}
\and

\IEEEauthorblockN{ Melike Erol-Kantarci,\IEEEmembership{ Senior Member, IEEE}}
\IEEEauthorblockA{\textit{School of Electrical Engineering and Computer Science} \\
\textit{University of Ottawa}\\
melike.erolkantarci@uottawa.ca}

\and
\IEEEauthorblockN{ Mahdi Hirab}
\IEEEauthorblockA{\textit{VMware Inc.}\\
mhirab@vmware.com}

\and
\IEEEauthorblockN{ Hoda Dehghan}
\IEEEauthorblockA{\textit{VMware Inc.}\\
hdehghan@vmware.com}

\and
\IEEEauthorblockN{Rodney Wilson,\IEEEmembership{ Senior Member, IEEE}}
\IEEEauthorblockA{\textit{Ciena Corp.} \\
rwilson@ciena.com}
\\\textcolor{white}{School of Electrical Engineering}

}
\maketitle
\begin{abstract}
The flexibility offered by reconfigurable wireless networks, provide new opportunities for various applications such as online AR/VR gaming, high-quality video streaming and autonomous vehicles, that desire high-bandwidth, reliable and low-latency communications. These applications come with very stringent Quality of Service (QoS) requirements and increase the burden over mobile networks. Currently, there is a huge spectrum scarcity due to the massive data explosion and this problem can be solved by helps of Reconfigurable Wireless Networks (RWNs) where nodes have reconfiguration and perception capabilities. Therefore, a necessity of AI-assisted algorithms for resource block allocation is observed. To tackle this challenge, in this paper, we propose an actor-critic learning-based scheduler for allocating resource blocks in a RWN. Various traffic types with different QoS levels are assigned to our agents to provide more realistic results. We also include mobility in our simulations to increase the dynamicity of networks. The proposed model is compared with another actor-critic model and with other traditional schedulers; proportional fair (PF) and Channel and QoS Aware (CQA) techniques. The proposed models are evaluated by considering the delay experienced by user equipment (UEs), successful transmissions and head-of-the-line delays. The results show that the proposed model noticeably outperforms other techniques in different aspects. 
\end{abstract}


\section{Introduction}
With the growing demand on social media platforms, Ultra-High-Definition (UHD) video and Virtual Reality (VR), Augmented Reality (AR) enabled applications, the ever-growing data traffic of Internet-of-Things (IoT), and the speedy advances in mobile devices and smartphones, have led to exponential growth in the traffic load over wireless networks \cite{b13}. Due to these new applications, availability of various traffic types, and unpredictability of physical channels, including fading, path loss, etc., maintaining quality of service (QoS) has become more challenging than ever before. In recent years, with the introduction of Open Radio Access Network (O-RAN), Artificial Intellience (AI) and Machine Learning (ML) have found applications in wireless networks, and Reconfigurable Wireless Networks (RWNs) have emerged. In RWNs, local networking nodes are controlled by groups of communicating nodes equipped with reconfigurable software, hardware, or protocols. Software reconfiguration is useful for updating, inclusion, and exclusion of tasks while hardware reconfiguration will enable manipulating physical infrastructure. AI and ML techniques can provide automation at a higher degree than before (further than self-organized networks (SON) concept of 3GPP) and manage the growing complexity of the RWNs \cite{b14}.  

Reinforcement learning (RL) is a machine learning technique that allows optimal control of systems by directing the system to a desired state by interacting with the environment and using feedback from the environment \cite{c1}. RL techniques are widely used in cellular networks with various use cases such as video flow optimization \cite{b16}, improving energy-efficiency in mobile networks \cite{b17}, and optimizing resource allocation \cite{b15}. A majority of RL-based methods used in wireless networks focus on Q-learning or Deep Q-learning. Although promising results are obtained using these techniques, they techniques offer a single level control. Instead, actor-critic learning which is a type of RL, can be implemented in multiple hierarchies and offer control from multiple points of views.  

In this work, we propose an Actor-Critic Learning approach \cite{c2}, to allocate resource blocks in a way that the communication reliability is enhanced and the required QoS by each UE is satisfied. In the proposed model, we formulate the choice of the number of resource blocks (RBs) and the location of them in the RBs’ map as a Markov Decision Process (MDP) \cite{c3}, and we solve this problem by using an actor-critic model. We consider channel quality, packets priorities, and the delay budget of each traffic stream in our reward function. We adopt two Advantage Actor-Critic (A2C) models \cite{c4}. The first technique solely schedules packets by giving priority to their scheduling delay budget (called as D-A2C) while second technique considers channel quality, delay budget, and packet types (called as CDPA-A2C). We evaluate the performance of the proposed models using NS3 \cite{c5} with fixed and mobile scenarios. Our results show that, in the fixed scenario, the proposed model can reduce the mean delay significantly with respect to proportional fair (PF) \cite{c6}, Channel and QoS Aware (CQA) \cite{b12} and D-A2C schedulers. Additionally, CDPA-A2C can increase the packet delivery rate in the mobile scenario up to 92\% and 53\% in comparison with PF and CQA.

The main contributions of this paper are:
\begin{itemize}


\item Proposing an actor-critic learning technique that can be implemented on disaggregated RAN functions and provide control in two levels.

\item Proposing a comprehensive reward function which takes care of channel quality, packet priorities, and the delay budget of each traffic stream.

\item Proposing two A2C models, where the first model solely schedules  packets by giving priority to their scheduling delay budget (called as D-A2C) and  the second technique considers channel quality, delay budget, and packet types (called as CDPA-A2C).


\end{itemize}

The rest of the paper is organized as follows. In Section II, we summarized the related works. In Section III, the system model is described. In Section IV, the proposed actor-critic resource block scheduler is explained. Numerical results and evaluation of the proposed model are presented in Section V, and finally, we conclude the paper in Section VI.

\section{Related Work}
Providing ubiquitous connectivity for various devices with different QoS requirements is one of the most challenging issues for mobile network operators \cite{c7}. This problem is amplified in future 5G applications with strict QoS requirements \cite{c8}. Additionally, in order to be capable of handling all the new immersive applications (which are known for their heterogeneous QoS properties), advanced techniques are required to maintain quality of experience (QoE) among the network’s entities. To this end, packet schedulers need to allow sharing the network bandwidth dynamically among UEs in a way that UEs achieve their target QoS. Many scheduling algorithms have been introduced previously, which employs QoS in their models. In \cite{b1}, a scheduler is proposed, which encapsulates different features of scheduling strategies for the downlink of cellular networks to guarantee multi-dimensional QoS for various radio channels and traffic types. However, most of the QoS-based schedulers are prioritizing some traffic types by ignoring the rest. For instance, in \cite{b2}, a prioritized traffic scheduler named by frame level scheduler (FLS) is introduced, which gives higher priority to real-time traffics in comparison with elastic traffics (such as HTTP or file transfer). Additionally, in \cite{b3} required activity detection (RADS) scheduler is proposed, which prioritized UEs based on the fairness and their packet delay. However, most of prioritizing schedulers are not capable of quickly reacting to the dynamics of cellular networks. Therefore, some traffic classes may have degradation in their QoS, while others can be over-provisioned.

RL-based models are also applied in different ways in order to optimize resource allocation in networks. In \cite{e3}, an RL-based scheduler is presented for resource allocation in a reliable vehicle-to-vehicle (V2V) network. The presented RL scheduler interacts with the environment frequently to learn and optimize resource allocation to the vehicles. In this work, it is assumed that the whole network structure is connected to a centralized scheduler. Additionally, in \cite{e4}, resource allocation and computation offloading in multi-channel multi-user mobile edge cloud (MEC) systems are evaluated. In this work, the authors presented a deep reinforcement network to jointly optimize the total delay and energy consumption of all UEs. Moreover, in \cite{e5}, an RL controller is implemented to schedule deadline-driven data transfers. In this paper, it is assumed that the requests will be sent to a central network controller where the flows with respect to their pacing rates can be scheduled. In \cite{e1,e2}, authors introduce a traffic predictor for network slices with a low complexity, based on a soft gated recurrent unit. They also use the traffic predictor to feed several deep learning models, which are trained offline to apply end-to-end reliable and dynamic resource allocation under dataset-dependent generalized service level agreement (SLA) constraints. The authors have considered resource bounds-based SLA and violation rate-based SLA in order to estimate the required resources in the network.

Traditional radio resource management (RRM) algorithms are not able to handle the stringent QoS requirements of users while adapting to fast varying conditions of RWNs. Recently, machine learning algorithms have been employed in schedulers to benefit from data in optimizing resource allocation, as opposed to using models \cite{b5,b6,b7,b8}. An ML-based RRM algorithm is proposed in \cite{b4} to estimate the required resources in the network for tackling traffic on-demand traffics over HTTP. ML has been used in resource allocation by considering various QoS combinations objectives such as packet loss \cite{b5}, delay \cite{b6}, and user fairness \cite{b7}. However, these models are defined for improving delay or enhancing throughput for Ultra-Reliable and Low-Latency Communications (URLLC) and throughput of enhanced Mobile Broadband (eMBB) UEs, while traffic types are considered homogeneous \cite{b8}. Hence, they ignored the effect of traffic types with various QoS requirements in their model. In comparison with previous works, we propose actor-critic learning for resource allocation which can be used at different levels of disaggregated RANs. We use a reward function that addresses channel quality, packet priorities, and the delay budget of each traffic stream. We evaluate our proposed scheme under various traffic types, interference levels and mobility. We provide detailed results on key performance indicators (KPIs) collected from NS3 simulator and integrated Open-AI Gym. 

\begin{figure*}[]
\centering
\includegraphics [width=.75\linewidth] {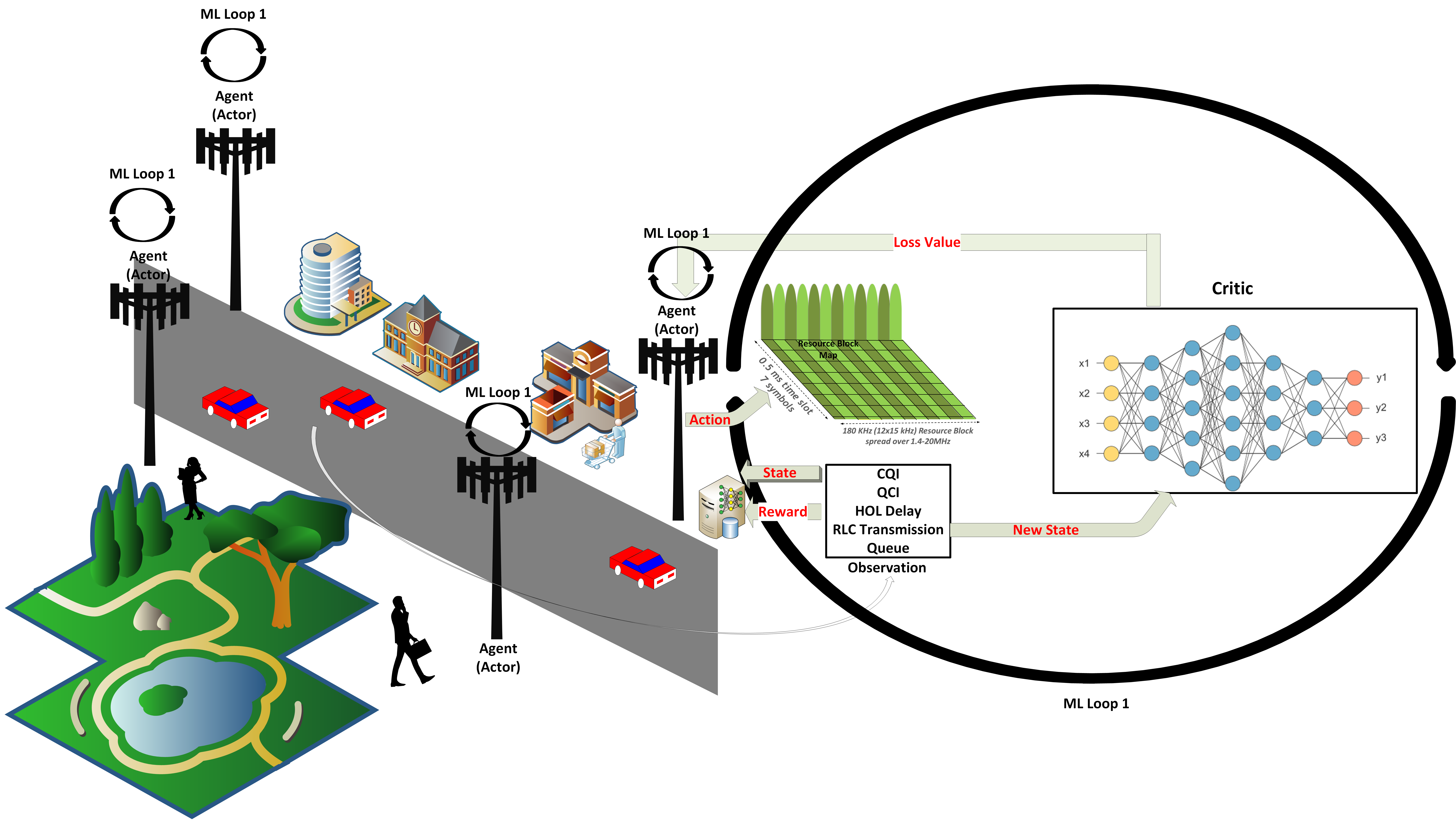} 
\caption{A Distributed RL-based RB Scheduler.}
\label{rl-network}
\end{figure*}
\section{System Model}
 We assume the overall downlink bandwidth is divided into the total number of available RBs, and each RB contains 12 contiguous subcarriers. Moreover, a resource block group (RBG) will be formed when consecutive RBs are grouped together. In order to reduce the number of state's in our reinforcement learning approach, we consider RBG as a unit for resource allocation in the frequency domain. The aim of the proposed actor-critic model is to assign RBGs by considering traffic types, their QoS requirement, and their priorities during each transmission time interval (TTI). Based on our system model, Base Stations (BS) are actors and each actor schedules the packets located in the transmission buffer of its associated UEs during each time interval such that, the amount of time that the packets stay in UEs buffers are reduced. The scheduling decision will be made every TTI, by considering the number of pending packets in the transmission buffers of active UEs. The overall delay experienced by a packet can be break down into three main factors as shown below:
\begin{equation}
    Packet_{Latency}=T_{HOL}+T_{tx}+T_{HARQ}
\end{equation}
where $T_{HOL}$ is the time duration that a packet waits in the queues to get a transmission opportunity (scheduling delay). HOL stands for head-of-the-line delay. $T_{tx}$ is communication delay, and $T_{HARQ}$ is a round-trip time which is required to retransmit a packet. $T_{tx}$ and $T_{HARQ}$ are basically based on the environment (path loss, shadowing, fading, etc.), UEs locations (propagation distance) and channel condition (noise, interference, etc.). In order to satisfy the packets with low latency requirements (e.g., URLLC UEs), the scheduler needs to handle those packets in UE buffers as soon as they arrive, thus, minimizing HOL. We also need to limit the number of HARQ to achieve lower delay during communication. However, limiting retransmissions can increase the packet drop rate and reduce the reliability in the network \cite{c9}. Low reliability can highly affect the UEs located at the edges. The proposed RL-based scheduler aims to address this trade off by enhancing reliability and meeting the required latency budget.

\section{Actor-Critic Scheduling Framework}
It is well-known that due to the scheduler's multi-dimensional and continuous state space, we can not enumerate the scheduling problem exhaustively  \cite{b10}. We can tackle this issue by employing RL and learning the proper scheduling rule. In actor-critic learning, policy model and value function are the two main parts of policy gradients. In order to reduce the gradient variance in our policy, we need to learn the value function for update and assist system policy during each time interval, and this is known as the Actor-Critic model. At the exploitation step, in order to make decisions, the learnt actor function is used. In our model, we aim to prioritize certain traffic types and reduce their HOL during scheduling decision. The obtained $M$ dimensional decision matrix is employed to schedule and prioritize the available traffic classes at each TTI. To do so, a neural network (NN) framework is employed to tackle the complexity and obtain an approximation for achieving the best possible prioritizing decision during each time interval. At the learning state, the weights of the neural network will be updated every TTI by considering the interactions occurring between the actor and the critic. Moreover, during the exploitation state, the value of the updated weights is saved, and the NN is tuned as a non-linear function. 

An RL agent basically tries to achieve higher value from each state $s_t$ and it can be obtained through the state-value ($V(s)$) and action-value functions $Q(s,a)$. Using the action-value function, we can estimate the output of action $a$ in state $s$ during time interval $t$, and the average expected output of state $s$ can be obtained by using the state-value function. In this work, instead of approximating both of action-value and state-value functions, we estimate only $V(s)$ by employing the Advantage Actor-Critic (A2C) model, which simplifies the learning process and reduces the number of required parameters. More specifically, advantage refer to a value which determines how much the performed action is better than the expected $V(s)$ ($A(s_t,a_t)=Q(s_t,a_t)-V(s_t)$). Moreover, the A2C model is a synchronous model and, with respect to asynchronous actor-critic (A3C) \cite{c10}, it provides better consistency among agents, making it suitable for disaggregated deployments.
 
The proposed A2C model contains two neural networks:
\begin{itemize}
\item A neural network in the critic for estimating the value function to criticize the actors’ actions during each state.
\item A neural network in the actor for approximating scheduling rules and prioritizing packet streams during each state. 
\end{itemize}
In the presented model, the critic is responsible for inspecting the actors’ actions and enhance their decisions at each time interval and its located at an edge cloud while the actor is at the BS. This flexible architecture is completed with recent O-RAN efforts around disaggregation of network functions. The high level perspective of the proposed model's architecture is presented in Fig.\ref{rl-network}.

In the following we present the actor-critic architecture step by step:
Actor: We employed actor as an agent to explore the required policy $\pi$ ($\theta$ is policy parameter) based on its observation ($O$) to obtain and apply the corresponding action ($A$).
\begin{equation}
\pi_{\theta }(O)=O\rightarrow A
\end{equation}
Therefore, the chosen action by an agent can be present as:
\begin{equation}
a=\pi_{\theta }(O),
\end{equation}
where $a\in A$. Actions are considered as choosing a proper resource block in the resource block map of each agent. Due to the discrete nature of actions, we employ softmax functions at the last layer (output) of actor to obtain the corresponding values of each actions. The summation of actions’ scores is equal to 1 and they are presented as probabilities of achieving a high reward value with respect to the chosen action. 


Critic: We employed the critic for obtaining the value function $V(O)$. During each time interval $t$, after the agent executed the chosen action by actor ($a_t$), it will send it to the critic along with the current observation ($O_t$). Then, the critic estimates the temporal difference (TD) by considering the next state ($O_{t+1}$) and the reward value ($R_t$) as follows:
\begin{equation}
\delta _t= R_t+\gamma V (O_{t+1})- (O_{t}).
\end{equation}

Here, $\delta _t$ is TD error for the action-value at time $t$, and $\gamma$ is a discount factor.  At the updating step, the least squares temporal difference (LSTD) need to be minimized to update the critic during each step:
\begin{equation}
V^*=arg \ \underset{V}{min}(\delta _t)^2,
\end{equation}
Here, the optimal value function is presented as $V^*$.
The actor can be updated by policy gradient which can be obtained by using the TD error as follow:
\begin{equation}
\bigtriangledown_{\theta } J(\theta)=E_{\pi_{\theta }}[\bigtriangledown_{\theta }log {\pi_{\theta }}(O,a)\delta _t],
\end{equation}
where, $\bigtriangledown_{\theta } J(\theta)$ is the gradient of the cost function with respect to $\theta$, and the value of action $a$ under the current policy is shown as $\pi_{\theta }(O,a)$. Then, the difference of parameters’ weights at the actor during time interval $t$ can be calculated as:
\begin{equation}
\Delta_{\theta_t}=\alpha \bigtriangledown_{\theta_t }log \pi_{\theta_t}(O_t,a_t)\delta _t,
\end{equation}
Here,  The  gradient is  estimated  per  time  step ($\bigtriangledown_{\theta_t }$)  and parameters will be updated in this gradient direction; also, the learning rate is defined as $\alpha$, which is between 0 and 1. Finally, the actor network, by using policy gradient can be updated as follows:
\begin{equation}
\theta_{t+1}=\theta_t+\alpha \bigtriangledown_{\theta_t }log \pi_{\theta_t}(O_t,a_t)\delta _t.
\end{equation}

Our main goal is to provide a channel, delay and priority aware actor-critic learning based scheduler (CDPA-A2C). Actors are located at BSs; therefore, during their observation, they can access the channel condition through Channel Quality Indicator (CQI) feedback of UEs from the control channels, and the amount of time packets will remain in the buffer in order to be scheduled. Moreover, actors can tune their future actions with respect to the received reward value at each iteration. In order to train the agents based on the network requirements, we need to include information about the channel condition, the load of transmission buffer, and the priority of packets into our reward function. The reward function in actor $i$ ($BS_i$) is designed as follows:
\begin{align}
R=\varphi R_1 +\tau R_2 +\lambda R_3,\\
R_1= max\left ( sgn\left ( cqi_k-\frac{\sum_{j=0}^{K}cqi_j}{K} \right ),0 \right ) \label{r1}\\
R_2= \left\{\begin{matrix}
1 & Packet_{URLLC}\\ 
0 & Otherwise
\end{matrix}\right.\label{r2}\\
R_3=sinc(\pi \left \lfloor \frac{Packet_{delay}}{Packet_{budget}} \right \rfloor) \label{r3}
\end{align}
Here, $cqi_k$ is the feedback send by $UE_k$ to agent $i$ at time interval $t$, $K$ is the total number of UEs associated with the agent $i$, $Packet_{URLLC}$ is an identifier for packets with a low delay budget, and it is associated to the $QCI$, $ Packet_{delay}$ is the HOL delay at RLC buffer and $Packet_{budget}$ is the maximum tolerable delay for the corresponding packet type, which is defined based on the packets’ types, $\varphi$, $\tau$ and $\lambda$ are scalar weights to control the priority among traffic types, maintaining delay budget and UEs condition (CQI feedback) \cite{e6,e7}. The reward function is tuned in a way that UEs received signal strength alongside packet delivery ratio be maximized while giving higher priority to the critical traffic load to increase QoS of URLLC users.

In this paper, to examine the effect of the proposed reward function, we defined two A2C models as follows:
\begin{itemize}
\item In the first one, the scheduler will schedule packets just based on packets' delay budget, and it is named as Delay aware A2C (D-A2C). In this model, the priority of packets (in our scenario URLLC packets) are ignored  by setting $\tau =0$ in eq.  (\ref{r2}).
\item In the second one, the scheduler takes all performance metrics into consideration by using eqs. (\ref{r1}),(\ref{r2}) and (\ref{r3}). The scheduler is called as CDPA-A2C. In this model, instead of giving priority to just one metric, the scheduler is equipped with a more complex model to be capable of handling RB allocation in different conditions. 

\end{itemize}

\section{Simulation Environment}

In this work, we implemented the proposed algorithms in ns3-gym \cite{c11}, which is a framework for connecting ns-3 with OpenAI Gym (a tool for integrating machine learning libraries) \cite{c12}. The neural networks of CDPA-A2C and D-A2C were implemented in Pytorch. In our simulations, three BSs are considered, and the number of UEs varies between 30 to 90 UEs, which are distributed randomly in the environment. Simulation results are based on 30 runs and each run contain 5000 iterations. To run the following model, we used a PC equipped with $Core^{TM}$ i7-8700 CPU and 32 GB of RAM. The simulation time depends on the number of assigned UEs and BSs, traffic types, and UEs' mobility. The simulation time can be varied between 30-90 minutes for simulating 5000 TTI based on the defined scenarios in this paper. In this work numerology zero is employed with 15 KHz subcarrier spacing, 12 subcarriers per resource block, and 14 symbols per subframe and our scheduling interval is set to 1 $ms$ \cite{c13}.

We deployed two scenarios. In the first scenario, we assume the number of UEs vary between 30 to 90 and UEs are not mobile. We distributed three traffic types (voice, video, and 
IP Multimedia Subsystem (IMS)) with different QoS requirements uniform randomly among UEs. The UE applications generate traffic with Poisson arrivals. Each traffic type has its own properties and different arrival time with respect to other packets.

In our simulations, when the amount of time a packet stays in a UE's buffer is lower than its delay budget, we consider it as a satisfied packet, and when this value is higher than the delay budget, it will be evaluated as an unsatisfied packet. In Table.I, we present the required QoS metrics for each traffic type considered in this paper, in detail.
\begin{table}[h]
\centering
\caption{The employed packet types and their properties \cite{b11}.}
\begin{tabular}{|l|l|l|l|l|}
\hline
QCI & \multicolumn{1}{c|}{\begin{tabular}[c]{@{}c@{}}Resource\\ Type\end{tabular}} & Priority & \begin{tabular}[c]{@{}l@{}}Packet\\ Delay\\ Budget\end{tabular} & \begin{tabular}[c]{@{}l@{}}Service \\  Example\end{tabular} \\ \hline
1   & GBR                                                                          & 2        & 100 ms                                                          & Voice                                                         \\ \hline
5   & Non-GBR                                                                      & 1        & 100 ms                                                          & IMS                                                           \\ \hline
6   & Non-GBR                                                                      & 6        & 300 ms                                                          & Video                                                         \\ \hline
75  & GBR                                                                          & 2.5      & 20 ms                                                           & V2X                                                           \\ \hline
\end{tabular}
\label{packets}
\end{table}

In the second scenario, we have 70-110 UEs in our vehicular network, which 10\% of them are vehicular UEs ($UE_v$) and the rest are fixed users ($UE_c$). The $UE_c$ requires a high capacity link, while $UE_v$ demands need to be satisfied by a low latency link. In this scenario, in addition to “Voice, Video, and IMS” we have Vehicle-to-everything “V2X” packets that are defined based on 5GAA standards \cite{c14}. Table II presents the assigned network parameters and neural network settings in our simulations. 
$\varphi$, $\tau$ and $\lambda$
\begin{table}[h]
\centering
\caption{Simulation parameters.}
\begin{tabular}{|c|c|}
\hline
Parameters                                                                       & Value                           \\ \hline
Number of neurons                                                                & 256 x 3 layers (Actor + Critic) \\ \hline
Scheduler algorithm                                                              & CDPA-A2C, D-A2C, PF, CQA                         \\ \hline
Number of BSs                                                                    & 3                               \\ \hline
Number of UEs                                                                    & 30-110                           \\ \hline
\begin{tabular}[c]{@{}c@{}}Maximum Traffic load per UE\\ (Downlink)\end{tabular} & 256 kbps                        \\ \hline
Traffic types                                                                    & Voice, Video, IMS, V2X          \\ \hline
Traffic stream per TTI                                                                    & 50          \\ \hline
D-A2C reward's weights                                                                  & \makecell{$\varphi = 1-\frac{packet_{delay}}{traffic_{delay-budget}}$,\\ $\tau=0$ and $\lambda=5$ }  \\ \hline
CDPA-A2C reward's weights                                                                  & \makecell{$\varphi = 1-\frac{packet_{delay}}{traffic_{delay-budget}}$,\\ $\tau=5$ and $\lambda=5$ }  \\ \hline
Discount factor                                                                  & 0.9                             \\ \hline
Actor learning-rate                                                              & 0.01                            \\ \hline
Critic learning-rate                                                             & 0.05                            \\ \hline
\end{tabular}
\end{table}

\begin{figure}[]
\centering
\includegraphics [width=.75\linewidth] {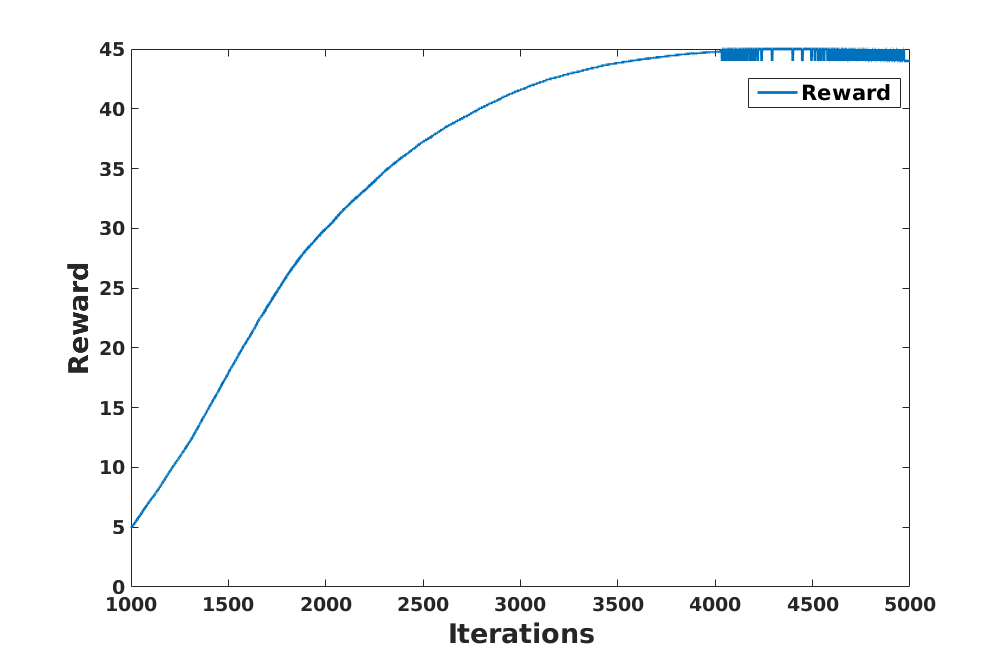} 
\caption{The convergence performance of the CDPA-A2C algorithm's reward function when the number of UEs are 90 while 10\% of them are mobile.}
\label{reward}
\end{figure}

Before presenting network performance results, in Fig.~\ref{reward} we show the convergence of the proposed reward. The figure shows the behavior of the reward function (eq. 9) when the number of UEs are 90 and 10\% of them are mobile. In this work, an epsilon-greedy policy is used, in which during the exploration phase actors perform their actions either by randomly or by choosing an RB with the highest weight assigned by the proposed actor-critic model.  As is shown in Fig. 2, the exploration phase will end after almost 3700 rounds, and the model will converge after that.

\begin{figure*}[]
  \begin{subfigure}{\textwidth}
  \includegraphics[width=.33\textwidth]{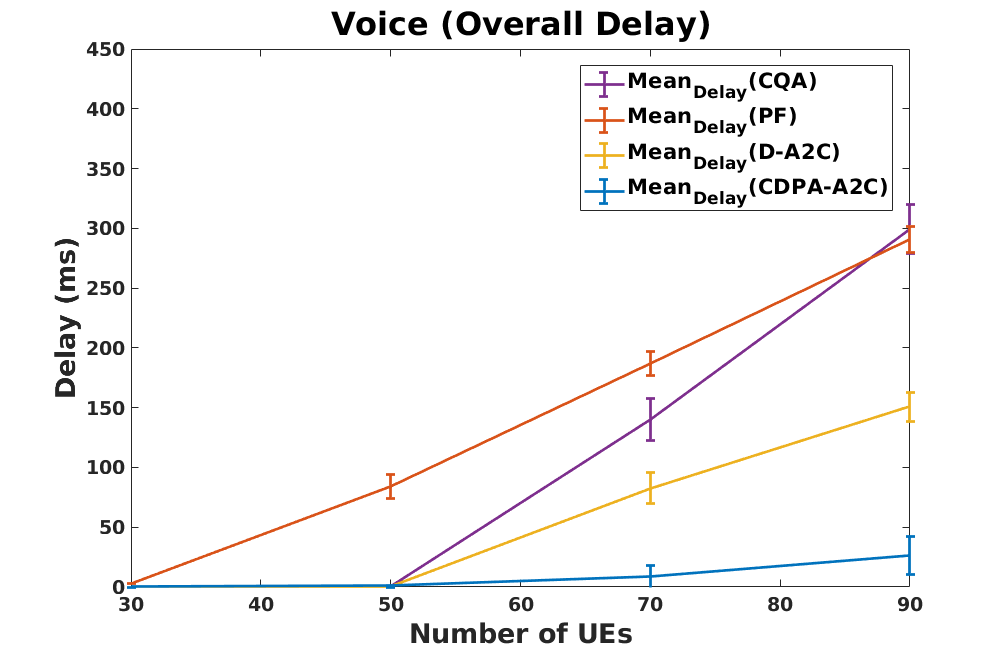}\hfill
  \includegraphics[width=.33\textwidth]{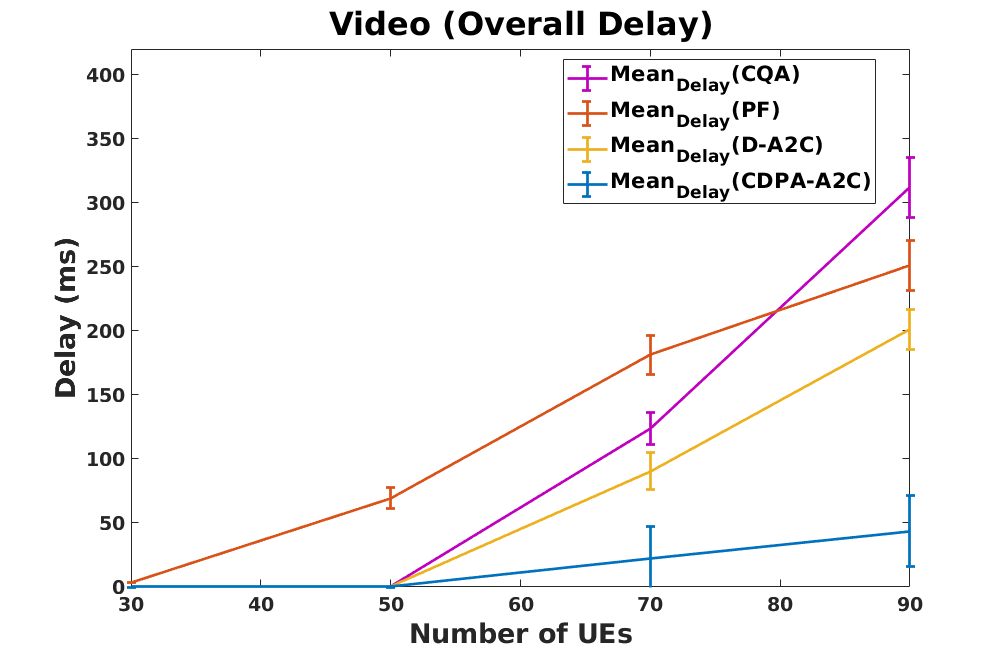}\hfill
  \includegraphics[width=.33\textwidth]{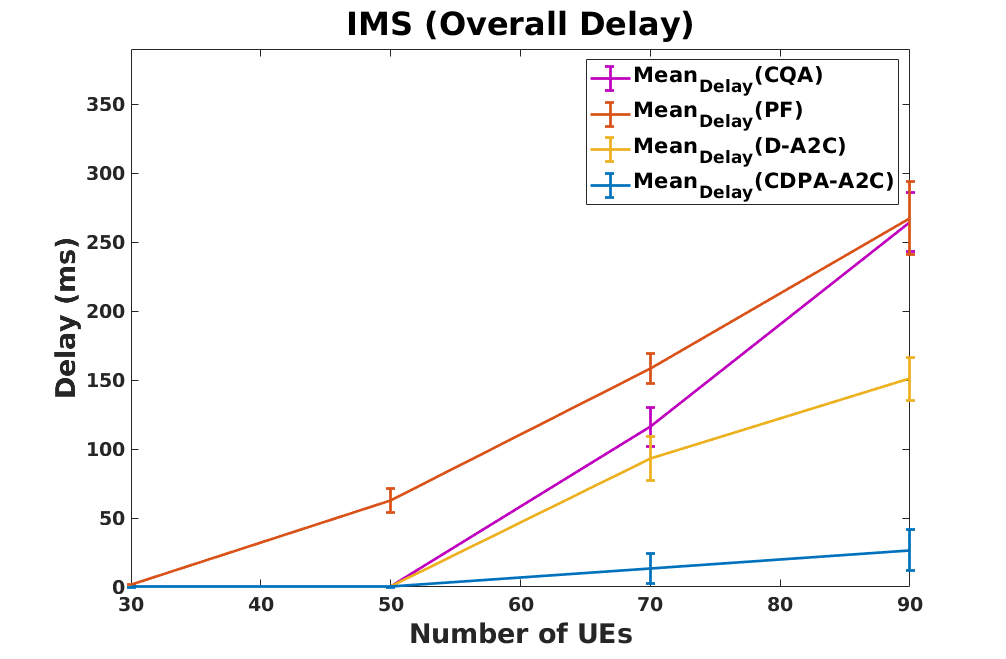}
  \end{subfigure}\par\medskip
  \caption{The HOL delay for different number of UEs (with no mobility).}
  \label{delays1}
\end{figure*}

\begin{figure*}[!h]
  \begin{subfigure}{\textwidth}
  \includegraphics[width=.33\textwidth]{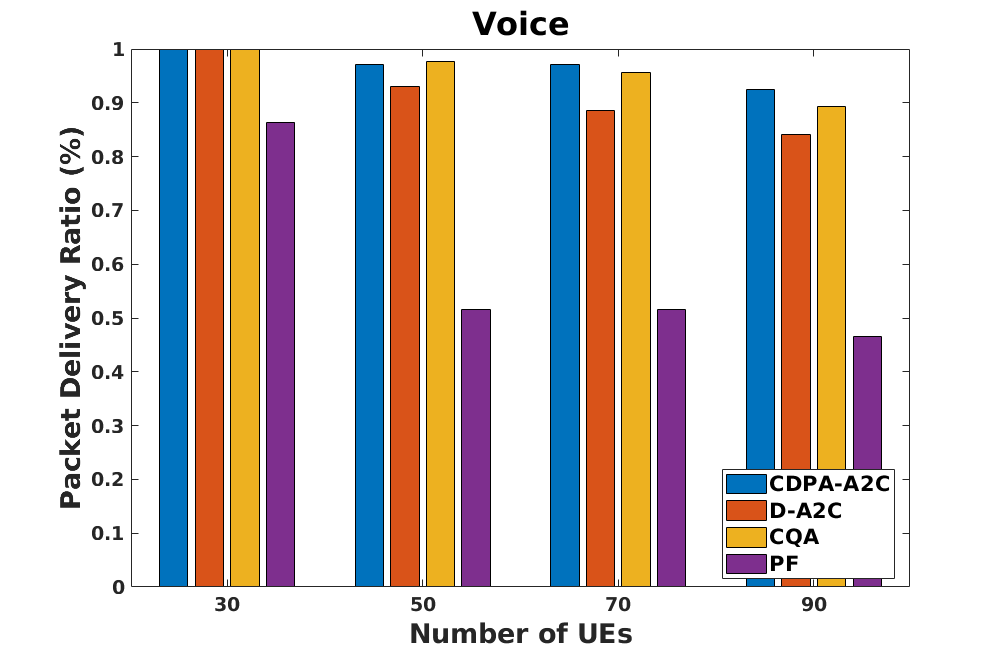}\hfill
  \includegraphics[width=.33\textwidth]{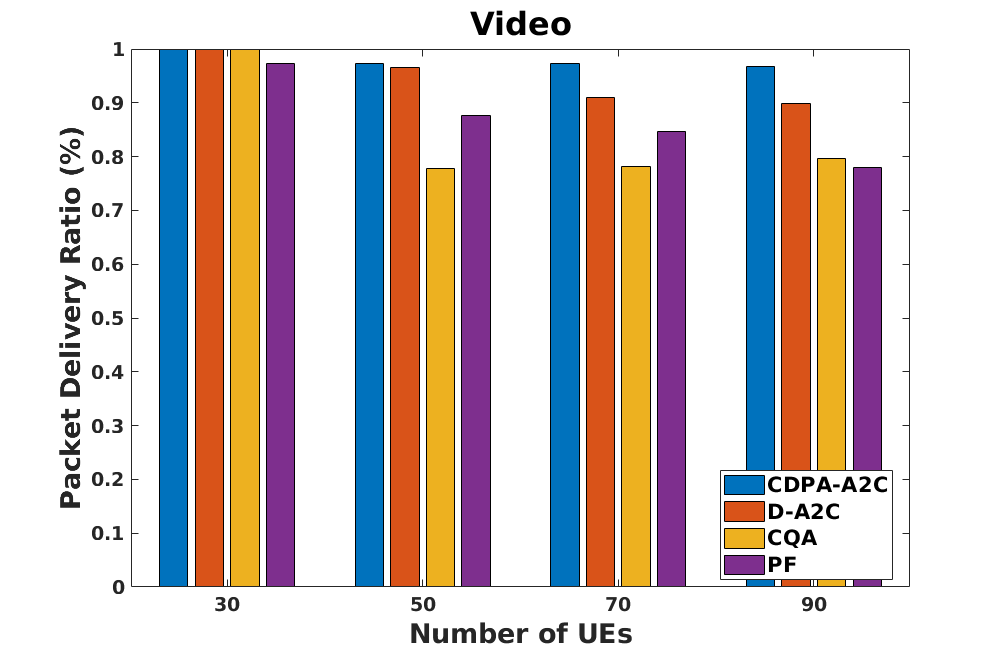}\hfill
  \includegraphics[width=.33\textwidth]{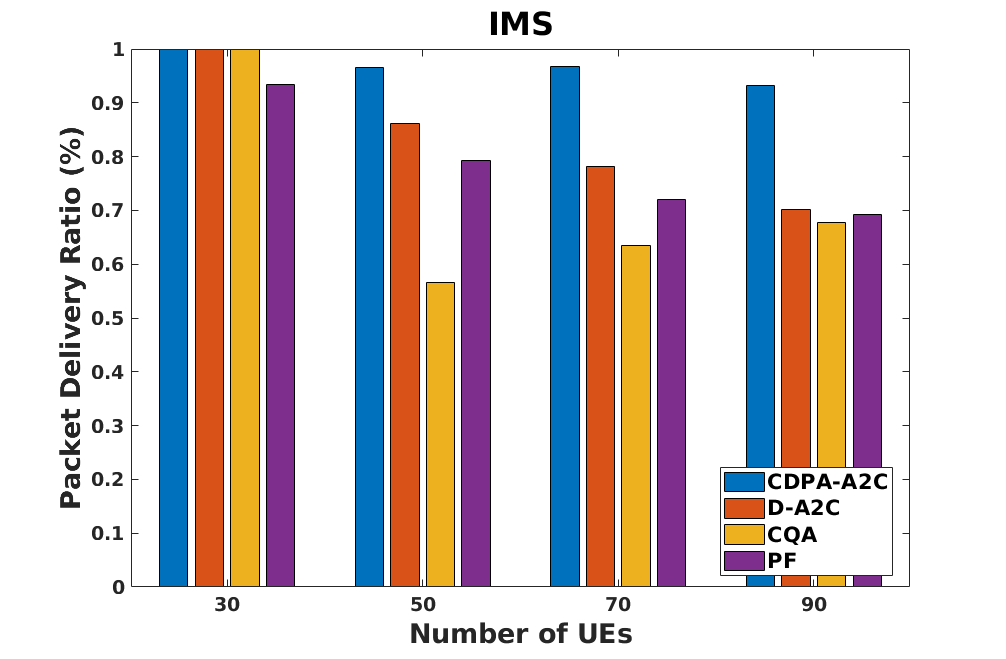}
  \end{subfigure}\par\medskip
  \caption{Packet delivery ratio (with no mobility).}
  \label{qoss1}
\end{figure*}

\subsection{Simulation Results with no mobility}

\begin{figure*}[]
  \begin{subfigure}{\textwidth}
  \includegraphics[width=.4\textwidth]{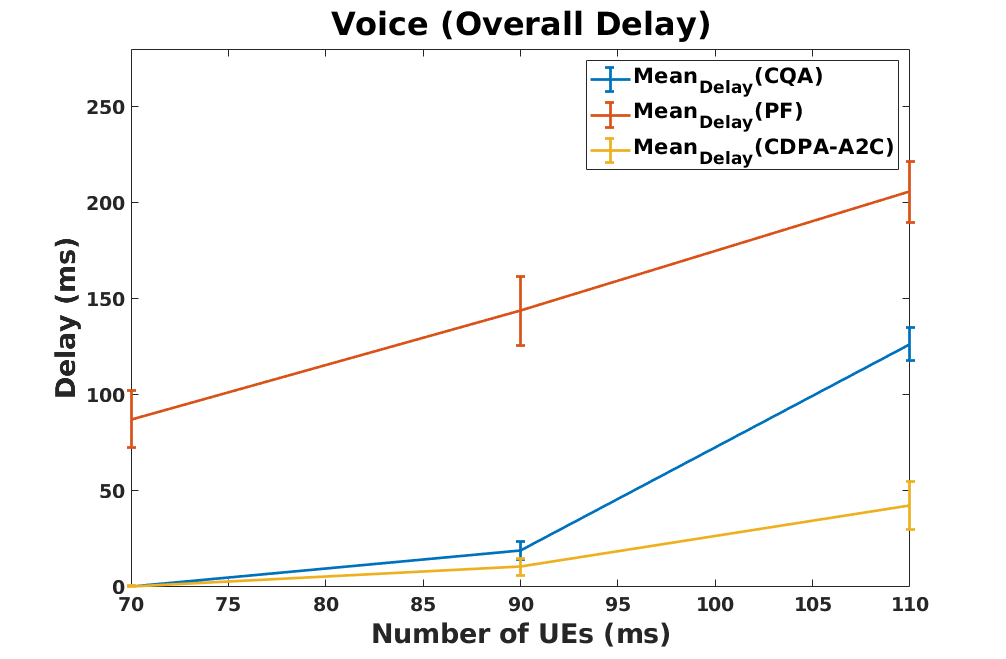}\hfill
  \includegraphics[width=.4\textwidth]{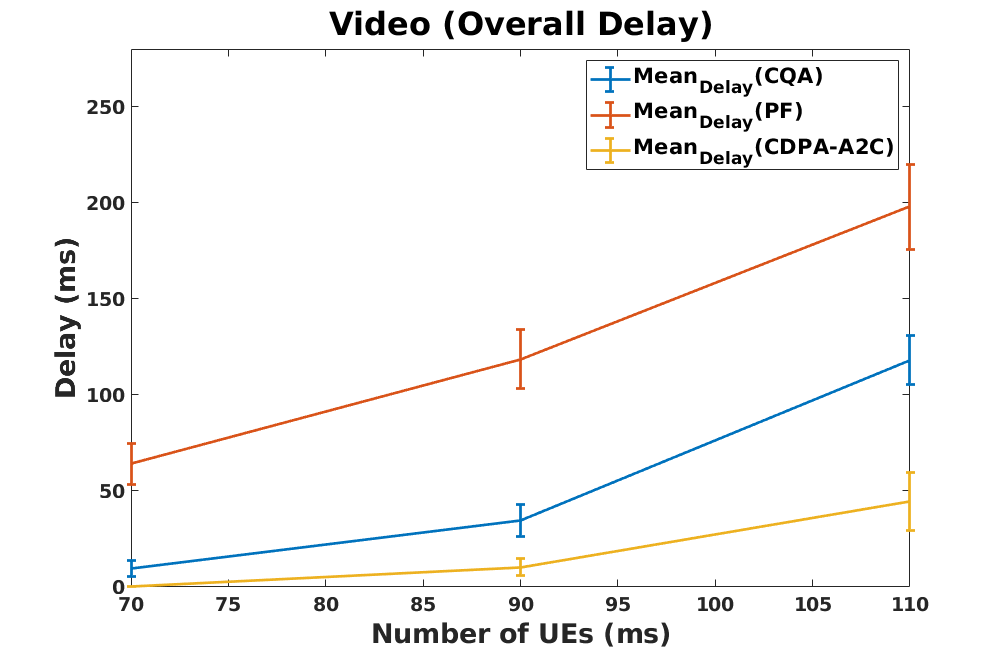}\hfill
  \includegraphics[width=.4\textwidth]{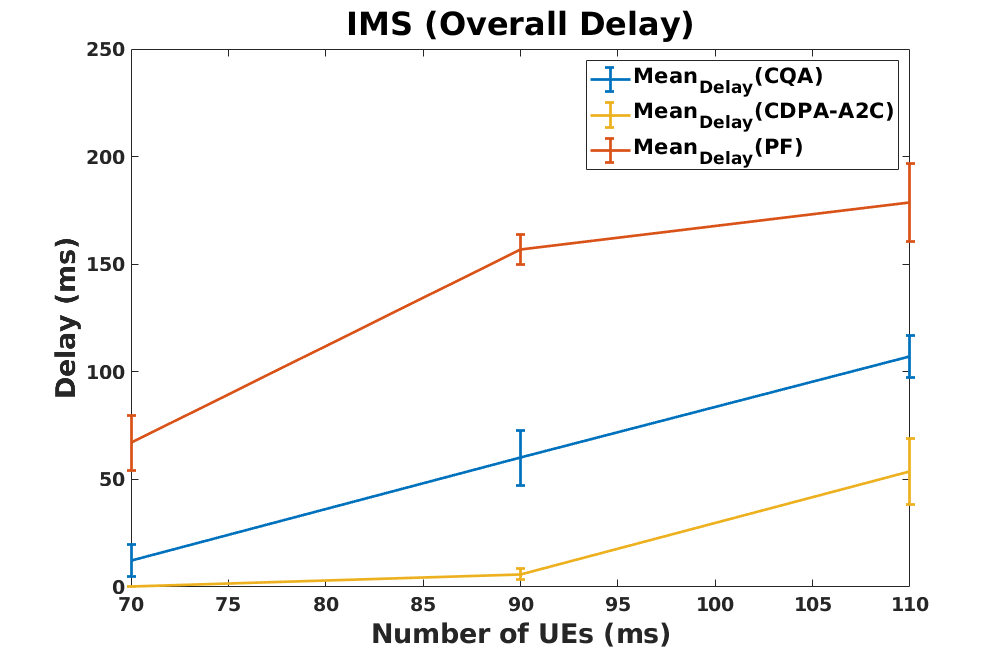}\hfill
  \includegraphics[width=.4\textwidth]{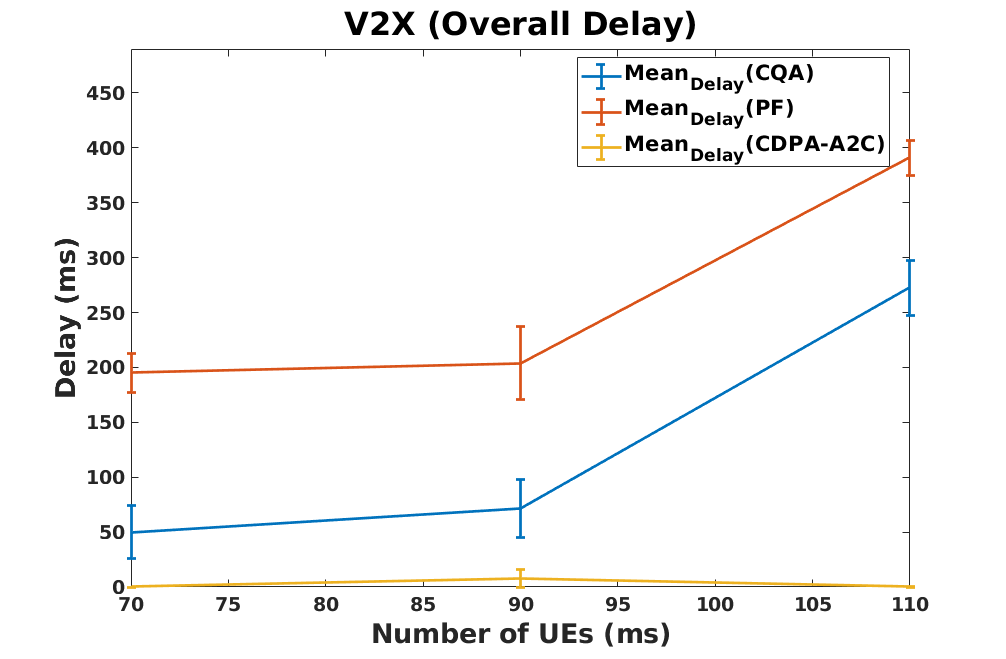}
  \end{subfigure}\par\medskip
  \caption{The HOL delay for different number of UEs (with mobility).}
  \label{delays2}
\end{figure*}

In this scenario, we assumed UEs are fixed, and we have three packet streams (voice, video, IMS) with different QoS requirements (delay budget, packet loss ratio, etc.). As it is mentioned previously, we considered two A2C models with different reward functions named by D-A2C and CDPA-A2C, respectively. In addition to these two models, we compare our results with the traditional Proportional Fair (PF) scheduler and Channel and QoS Aware (CQA) scheduler as described in \cite{b12}.

In Fig.~\ref{delays1}, we present the mean delay by considering various traffic types and varying number of UEs. Although, when the number of UEs below 50 (in case of CQA scheduler when the number of UEs are below 70), PF and CQA  can satisfy the required delay budget for all types of traffic, as we increase the number of UEs the mean HOL delay will be significantly increased, and it becomes higher than the target delay for voice and IMS traffic types. In cases with higher number of UEs, D-A2C results are better than the traditional PF and CQA schedulers, while employing a more comprehensive reward function such as CDPA-A2C can remarkably enhance network performance. In general, The CDPA-A2C scheduler can reduce the mean delay in comparison with D-A2C, PF and CQA schedulers up to 100 ms, 225 ms and 375 ms, respectively.


In Fig.~\ref{qoss1}, we present the packet delivery ratio (the ratio of the packets which satisfy the predefined delay budget). As it is shown, the packet delivery ratio of CDPA-A2C and D-A2C can be up to 117\% and 73\% higher than PF and CQA, by considering various numbers of UEs for different traffic types. Moreover, CDPA-A2C can considerably enhance the packet delivery ratio in comparison to D-A2C (up to 63\% for IMS packets). Note that, we assume all applications are sensitive to delay and will consider packets that are beyond service targets as undelivered.



\begin{figure*}[]
  \begin{subfigure}{\linewidth}
  \includegraphics[width=.4\linewidth]{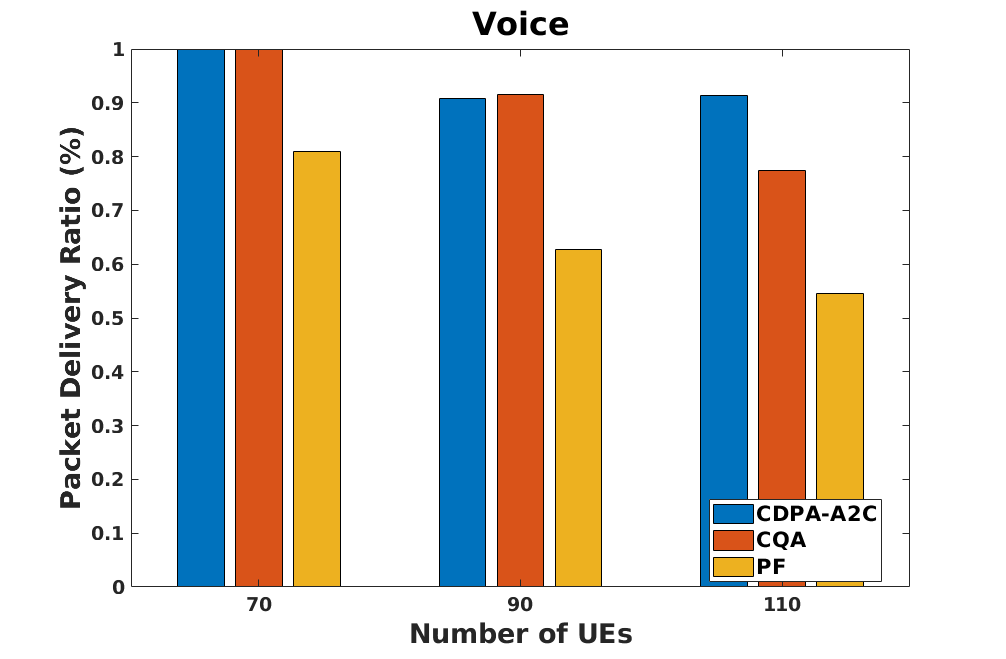}\hfill
  \includegraphics[width=.4\linewidth]{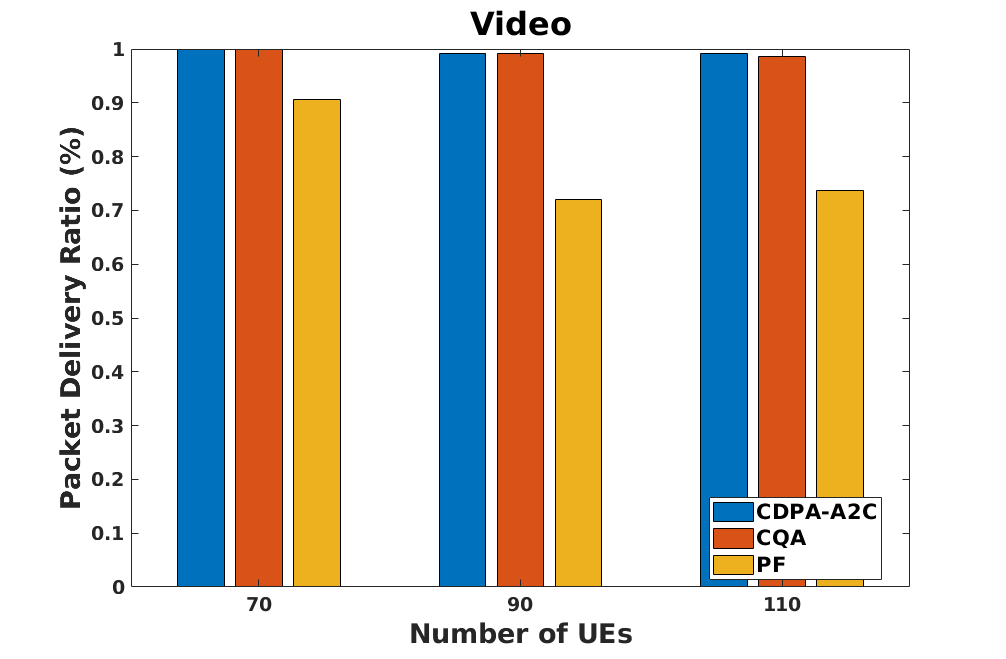}\hfill
  \includegraphics[width=.4\linewidth]{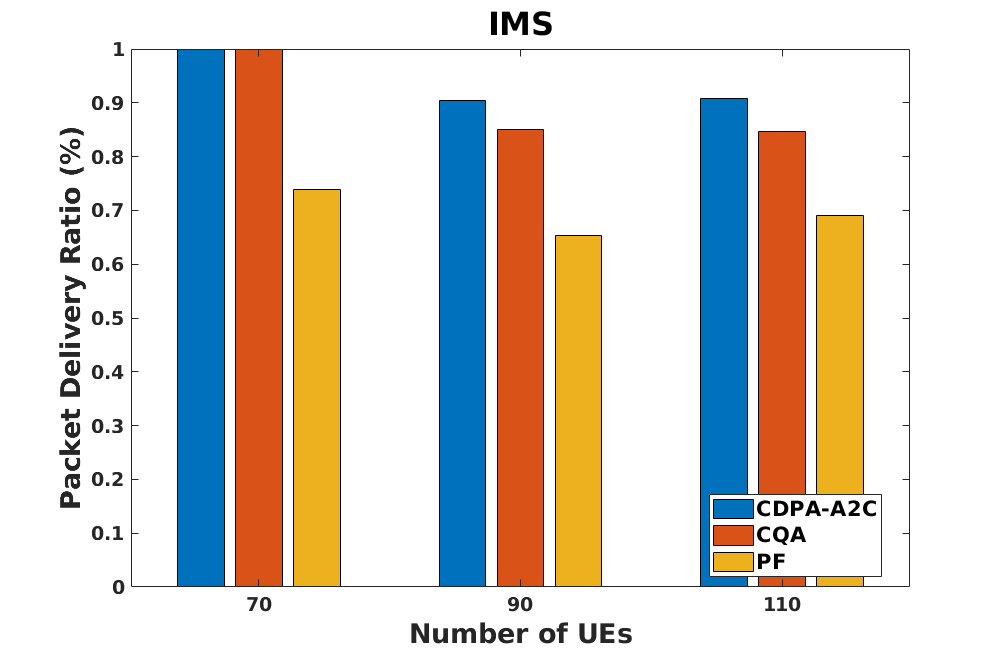}\hfill
  \includegraphics[width=.4\linewidth]{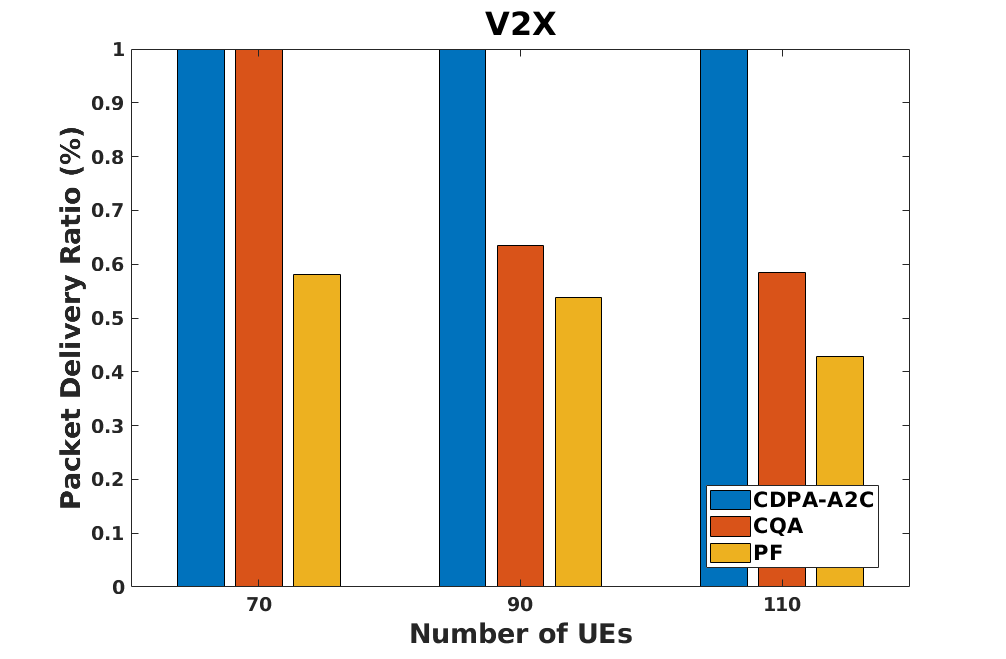}
  \caption{The effect of UEs' number on packet delivery ratio when the ratio of mobile UEs is fixed to 10\%.}
  \label{qoss2a}
  \end{subfigure}\par\medskip
    \begin{subfigure}{\linewidth}
  \includegraphics[width=.4\linewidth]{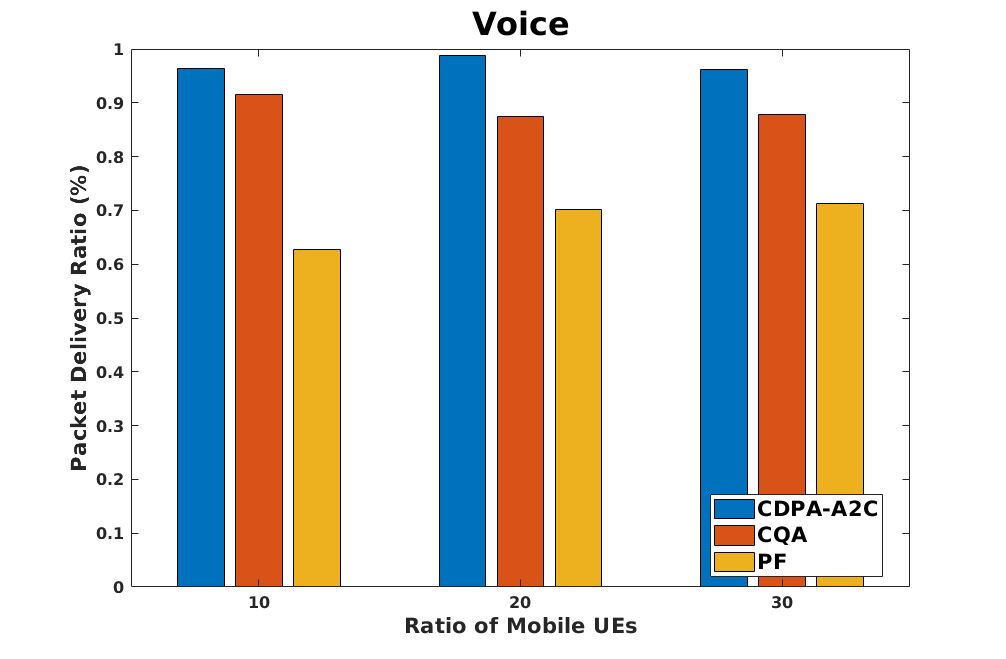}\hfill
  \includegraphics[width=.4\linewidth]{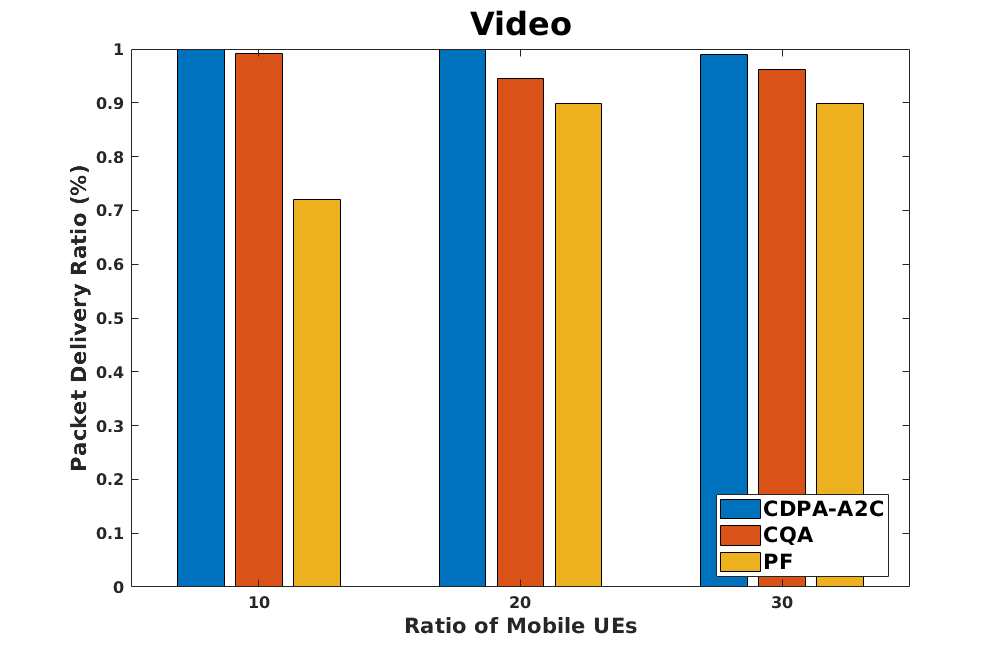}\hfill
  \includegraphics[width=.4\linewidth]{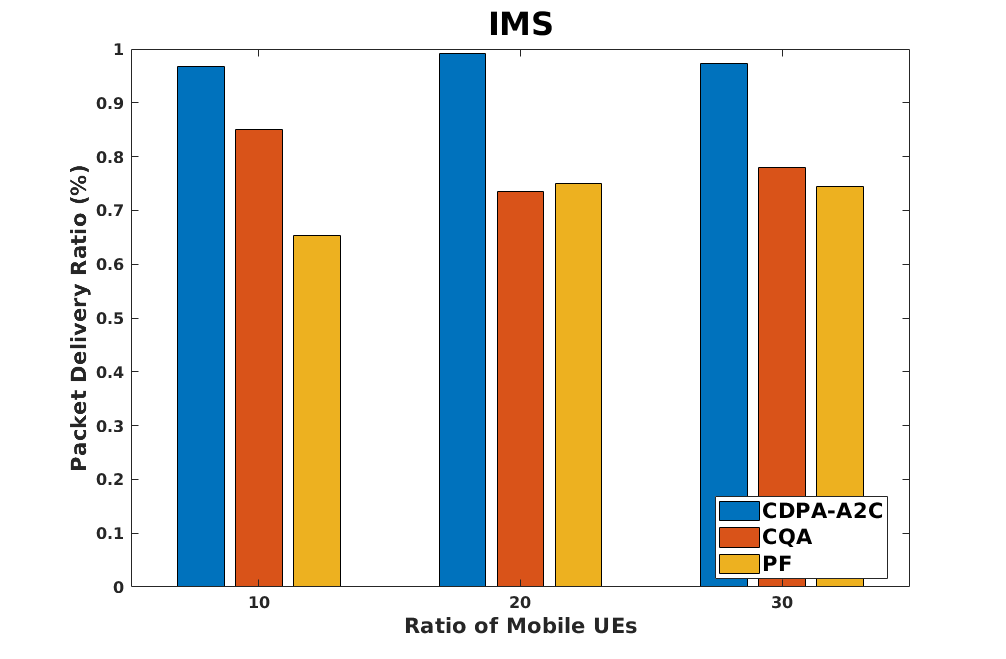}\hfill
  \includegraphics[width=.4\linewidth]{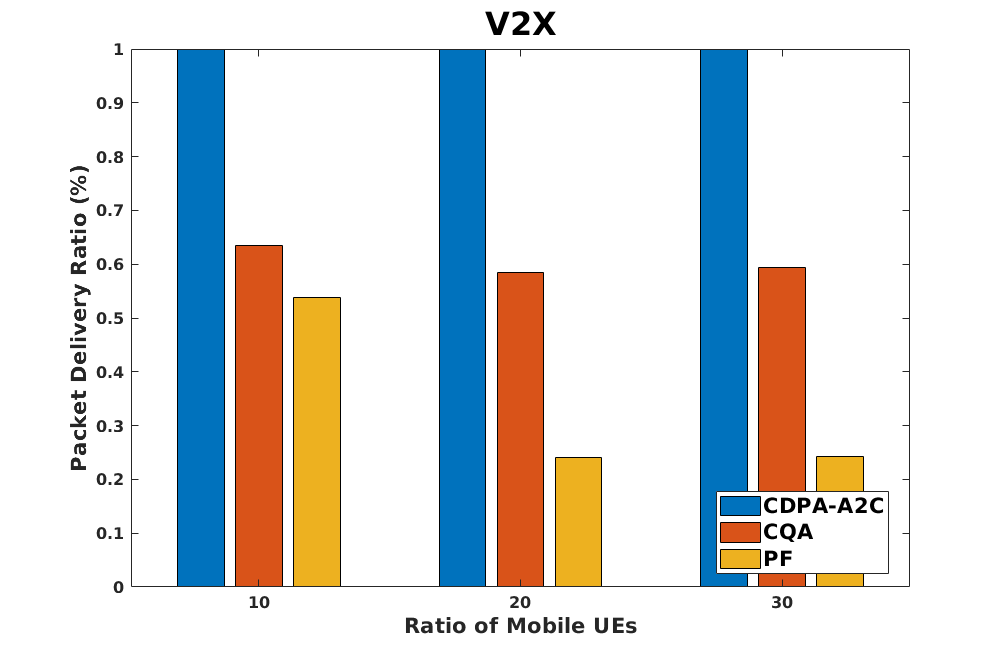}
  \caption{The effect of percentage of mobile UEs on packet delivery ratio when the total number of UEs is 90.}
  \label{qoss2b}
  \end{subfigure}\par\medskip
  \caption{Packet delivery ratio (with mobility).}
  \label{qoss2}
\end{figure*}

\subsection{Simulation Results with mobility}
In the second scenario, we consider a case when 10\% of UEs are mobile. We include V2X traffic based on 5GAA standards in addition to other traffic types used in the previous simulations. Due to the better performance of CDPA-A2C, we omit the D-A2C from second scenario. 

In Fig.~\ref{delays2}, we evaluate the mean HOL delay for various numbers of UEs (70-110) when we have mobile UEs in the network. As we can see, although, PF and CQA schedulers can maintain the mean HOL delay below the delay budget, by increasing the number of UEs, the mean HOL delay will increase dramatically. Moreover, due to the high sensitivity of V2X packets to delay (V2X delay budget = 20 $ms$), PF and CQA can not satisfy V2X packets in any of these cases. However, the proposed CDPA-A2C model, by scheduling packets on time and preventing the congestion in the UEs buffer, can provide a lower delay for the presented traffic types with respect to PF and CQA. 

In Fig.~\ref{qoss2}, we present the packet delivery ratio (packets that can satisfy delay target) of the proposed model with respect to other schedulers. In Fig.~\ref{qoss2a}, we evaluate the effect of increasing the number of UEs over the packet delivery ratio when 10\% of UEs are mobile. As shown, the packet delivery ratio of CDPA-A2C for different traffic types and numbers of UEs up to 92\% and 53\% higher than PF and CQA, respectively. Although CQA has a good performance in the delivery of Voice, Video, and IMS packets, CDPA-A2C can considerably enhance the packet delivery ratio for delay-sensitive traffics such as V2X packets in comparison with CQA. We also examined the effect of increasing the mobile UEs' ratio, on the packet delivery ratio, when the total number of UEs is fixed to 90, in Fig.~\ref{qoss2b}. As we can see, the proposed model can significantly enhance the packet delivery ratio for different traffic types with respect to PF and CQA. Moreover, by increasing the density of mobile UEs, the packet loss ratio of V2X packets when CQA and PF schedulers are employed, will be increased up to 40\% and 75\%, respectively. Therefore, the CDPA-A2C scheduler can noticeably enhance QoS by reducing the HOL delay and increasing the packet delivery ratio in comparison with PF and CQA schedulers. 




\section{Conclusion}
In this paper, we propose two actor-critic learning-based schedulers, namely Delay-aware actor-critic (D-A2C) and channel, delay and priority-aware actor-critic (CDPA-A2C) techniques, for reconfigurable wireless networks, where the network is capable to autonomously learn and adapt itself to the wireless environment's dynamicity for optimizing the utility of the network resources. Reconfigurable wireless networks play a vital role in network automation. Applying machine learning algorithms can be an appropriate candidate for making versatile models and making future networks smarter. Here, the proposed model is constructed based on two neural network models (actor and critic). Actors are employed to apply actions (RB allocation); simultaneously, the critic is used to monitor agents' actions and tune their behaviors in the following states to make the convergence faster and optimize the model. The proposed comprehensive model (CDPA-A2C), in addition of considering channel condition and delay budget of each packet, prioritizes the received packets by considering their types and their QoS requirements. We include fixed and mobile scenarios to evaluate the performance of the proposed schemes. We compare the learning-based schemes to two well-known algorithms: the traditional proportional fair and a QoS-aware algorithm CQA. Our results show that CDPA-A2C significantly reduces the mean delay with respect to PF and D-A2C schedulers. Additionally, CDPA-A2C can increase the packet delivery rate in the mobile scenario up to 92\% and 53\% in comparison with PF and CQA, respectively.

\section{Acknowledgement}
This work is supported by Ontario Centers of Excellence (OCE) 5G ENCQOR program.

\bibliographystyle{ieeetr}


\bibliography{references}

\begin{thebibliography}{10}

\bibitem{b13}
J.~{Navarro-Ortiz}, P.~{Romero-Diaz}, S.~{Sendra}, P.~{Ameigeiras}, J.~J.
  {Ramos-Munoz}, and J.~M. {Lopez-Soler}, ``{A Survey on 5G Usage Scenarios and
  Traffic Models},'' {\em IEEE Communications Surveys Tutorials}, vol.~22,
  no.~2, pp.~905--929, 2020.

\bibitem{b14}
M.~Polese, R.~Jana, V.~Kounev, K.~Zhang, S.~Deb, and M.~Zorzi, ``{Machine
  learning at the edge: A data-driven architecture with applications to {5G}
  cellular networks},'' {\em IEEE Transactions on Mobile Computing}, 2020.

\bibitem{c1}
N.~C. Luong, D.~T. Hoang, S.~Gong, D.~Niyato, P.~Wang, Y.-C. Liang, and D.~I.
  Kim, ``{Applications of deep reinforcement learning in communications and
  networking: A survey},'' {\em IEEE Communications Surveys \& Tutorials},
  vol.~21, no.~4, pp.~3133--3174, 2019.

\bibitem{b16}
M.~Polese, R.~Jana, V.~Kounev, K.~Zhang, S.~Deb, and M.~Zorzi, ``{Machine
  learning at the edge: A data-driven architecture with applications to 5G
  cellular networks},'' {\em IEEE Transactions on Mobile Computing}, 2020.

\bibitem{b17}
R.~Li, Z.~Zhao, X.~Zhou, G.~Ding, Y.~Chen, Z.~Wang, and H.~Zhang,
  ``{Intelligent 5G: When cellular networks meet artificial intelligence},''
  {\em IEEE Wireless communications}, vol.~24, no.~5, pp.~175--183, 2017.

\bibitem{b15}
S.~Chinchali, P.~Hu, T.~Chu, M.~Sharma, M.~Bansal, R.~Misra, M.~Pavone, and
  S.~Katti, ``{Cellular network traffic scheduling with deep reinforcement
  learning},'' in {\em Proceedings of the AAAI Conference on Artificial
  Intelligence}, vol.~32, 2018.

\bibitem{c2}
Y.~Wei, F.~R. Yu, M.~Song, and Z.~Han, ``{User scheduling and resource
  allocation in HetNets with hybrid energy supply: An actor-critic
  reinforcement learning approach},'' {\em IEEE Transactions on Wireless
  Communications}, vol.~17, no.~1, pp.~680--692, 2017.

\bibitem{c3}
C.~C. White, ``{A survey of solution techniques for the partially observed
  Markov decision process},'' {\em Annals of Operations Research}, vol.~32,
  no.~1, pp.~215--230, 1991.

\bibitem{c4}
O.~Nachum, M.~Norouzi, K.~Xu, and D.~Schuurmans, ``{Bridging the gap between
  value and policy based reinforcement learning},'' in {\em Advances in Neural
  Information Processing Systems}, pp.~2775--2785, 2017.

\bibitem{c5}
A.~K. Saluja, S.~A. Dargad, and K.~Mistry, ``{A Detailed Analogy of Network
  Simulators—NS1, NS2, NS3 and NS4},'' {\em Int. J. Future Revolut. Comput.
  Sci. Commun. Eng}, vol.~3, pp.~291--295, 2017.

\bibitem{c6}
M.~T. Kawser, H.~Farid, A.~R. Hasin, A.~M. Sadik, and I.~K. Razu,
  ``{Performance comparison between round robin and proportional fair
  scheduling methods for LTE},'' {\em International Journal of Information and
  Electronics Engineering}, vol.~2, no.~5, pp.~678--681, 2012.

\bibitem{b12}
B.~Bojovic and N.~Baldo, ``{A new channel and QoS aware scheduler to enhance
  the capacity of voice over LTE systems},'' in {\em 2014 IEEE 11th
  International Multi-Conference on Systems, Signals \& Devices (SSD14)},
  pp.~1--6, IEEE, 2014.

\bibitem{c7}
M.~F. Audah, T.~S. Chin, Y.~Zulfadzli, C.~K. Lee, and K.~Rizaluddin, ``{Towards
  Efficient and Scalable Machine Learning-Based QoS Traffic Classification in
  Software-Defined Network},'' in {\em International Conference on Mobile Web
  and Intelligent Information Systems}, pp.~217--229, Springer, 2019.

\bibitem{c8}
M.~A. Habibi, M.~Nasimi, B.~Han, and H.~D. Schotten, ``{A comprehensive survey
  of RAN architectures toward 5G mobile communication system},'' {\em IEEE
  Access}, vol.~7, pp.~70371--70421, 2019.

\bibitem{b1}
S.~Abedi, ``{Efficient radio resource management for wireless multimedia
  communications: a multidimensional QoS-based packet scheduler},'' {\em IEEE
  Transactions on Wireless Communications}, vol.~4, no.~6, pp.~2811--2822,
  2005.

\bibitem{b2}
G.~Piro, L.~A. Grieco, G.~Boggia, R.~Fortuna, and P.~Camarda, ``{Two-level
  downlink scheduling for real-time multimedia services in LTE networks},''
  {\em IEEE Transactions on Multimedia}, vol.~13, no.~5, pp.~1052--1065, 2011.

\bibitem{b3}
G.~Monghal, D.~Laselva, P.-H. Michaelsen, and J.~Wigard, ``{Dynamic packet
  scheduling for traffic mixes of best effort and VoIP users in E-UTRAN
  downlink},'' in {\em 2010 IEEE 71st Vehicular Technology Conference},
  pp.~1--5, IEEE, 2010.

\bibitem{e3}
T.~{\c{S}}ahin, R.~Khalili, M.~Boban, and A.~Wolisz, ``{Reinforcement learning
  scheduler for vehicle-to-vehicle communications outside coverage},'' in {\em
  2018 IEEE Vehicular Networking Conference (VNC)}, pp.~1--8, IEEE, 2018.

\bibitem{e4}
S.~Nath, Y.~Li, J.~Wu, and P.~Fan, ``{Multi-user Multi-channel Computation
  Offloading and Resource Allocation for Mobile Edge Computing},'' in {\em ICC
  2020-2020 IEEE International Conference on Communications (ICC)}, pp.~1--6,
  IEEE, 2020.

\bibitem{e5}
G.~R. Ghosal, D.~Ghosal, A.~Sim, A.~V. Thakur, and K.~Wu, ``{A Deep
  Deterministic Policy Gradient Based Network Scheduler For Deadline-Driven
  Data Transfers},'' in {\em 2020 IFIP Networking Conference (Networking)},
  pp.~253--261, IEEE, 2020.

\bibitem{e1}
H.~Chergui and C.~Verikoukis, ``{Offline SLA-constrained deep learning for 5G
  networks reliable and dynamic end-to-end slicing},'' {\em IEEE Journal on
  Selected Areas in Communications}, vol.~38, no.~2, pp.~350--360, 2019.

\bibitem{e2}
H.~Chergui and C.~Verikoukis, ``{Big Data for 5G Intelligent Network Slicing
  Management},'' {\em IEEE Network}, vol.~34, no.~4, pp.~56--61, 2020.

\bibitem{b5}
I.-S. Com{\c{s}}a, S.~Zhang, M.~E. Aydin, P.~Kuonen, Y.~Lu, R.~Trestian, and
  G.~Ghinea, ``{Towards 5G: A reinforcement learning-based scheduling solution
  for data traffic management},'' {\em IEEE Transactions on Network and Service
  Management}, vol.~15, no.~4, pp.~1661--1675, 2018.

\bibitem{b6}
I.-S. Comșa, S.~Zhang, M.~Aydin, P.~Kuonen, R.~Trestian, and G.~Ghinea, ``{A
  comparison of reinforcement learning algorithms in fairness-oriented OFDMA
  schedulers},'' {\em Information}, vol.~10, no.~10, p.~315, 2019.

\bibitem{b7}
M.~Elsayed and M.~Erol-Kantarci, ``{AI-enabled radio resource allocation in 5G
  for URLLC and eMBB users},'' in {\em 2019 IEEE 2nd 5G World Forum (5GWF)},
  pp.~590--595, IEEE, 2019.

\bibitem{b8}
M.~Mohammadi and A.~Al-Fuqaha, ``{Enabling cognitive smart cities using big
  data and machine learning: Approaches and challenges},'' {\em IEEE
  Communications Magazine}, vol.~56, no.~2, pp.~94--101, 2018.

\bibitem{b4}
A.~Martin, J.~Ega{\~n}a, J.~Fl{\'o}rez, J.~Montalb{\'a}n, I.~G. Olaizola,
  M.~Quartulli, R.~Viola, and M.~Zorrilla, ``{Network resource allocation
  system for QoE-aware delivery of media services in 5G networks},'' {\em IEEE
  Transactions on Broadcasting}, vol.~64, no.~2, pp.~561--574, 2018.

\bibitem{c9}
X.~Du, Y.~Sun, N.~B. Shroff, and A.~Sabharwal, ``{Balancing Queueing and
  Retransmission: Latency-Optimal Massive MIMO Design},'' {\em IEEE
  Transactions on Wireless Communications}, vol.~19, no.~4, pp.~2293--2307,
  2020.

\bibitem{b10}
I.-S. Com{\c{s}}a, G.-M. Muntean, and R.~Trestian, ``{An Innovative
  Machine-Learning-Based Scheduling Solution for Improving Live UHD Video
  Streaming Quality in Highly Dynamic Network Environments},'' {\em IEEE
  Transactions on Broadcasting}, 2020.

\bibitem{c10}
M.~Sewak, ``{Actor-Critic Models and the A3C},'' in {\em Deep Reinforcement
  Learning}, pp.~141--152, Springer, 2019.

\bibitem{e6}
M.~Elsayed and M.~Erol-Kantarci, ``Learning-based resource allocation for
  data-intensive and immersive tactile applications,'' in {\em 2018 IEEE 5G
  World Forum (5GWF)}, pp.~278--283, IEEE, 2018.

\bibitem{e7}
M.~Elsayed, M.~Erol-Kantarci, B.~Kantarci, L.~Wu, and J.~Li, ``Low-latency
  communications for community resilience microgrids: A reinforcement learning
  approach,'' {\em IEEE Transactions on Smart Grid}, vol.~11, no.~2,
  pp.~1091--1099, 2019.

\bibitem{c11}
P.~Gaw{\l}owicz and A.~Zubow, ``{NS-3 meets openai gym: The playground for
  machine learning in networking research},'' in {\em Proceedings of the 22nd
  International ACM Conference on Modeling, Analysis and Simulation of Wireless
  and Mobile Systems}, pp.~113--120, 2019.

\bibitem{c12}
G.~Brockman, V.~Cheung, L.~Pettersson, J.~Schneider, J.~Schulman, J.~Tang, and
  W.~Zaremba, ``Openai gym,'' {\em arXiv preprint arXiv:1606.01540}, 2016.

\bibitem{c13}
J.~Vihri{\"a}l{\"a}, A.~A. Zaidi, V.~Venkatasubramanian, N.~He, E.~Tiirola,
  J.~Medbo, E.~L{\"a}hetkangas, K.~Werner, K.~Pajukoski, A.~Cedergren, {\em
  et~al.}, ``{Numerology and frame structure for 5G radio access},'' in {\em
  2016 IEEE 27th annual international symposium on personal, indoor, and mobile
  radio communications (PIMRC)}, pp.~1--5, IEEE, 2016.

\bibitem{b11}
``{Table 6.1.7-A: Standardized QCI characteristics from 3GPP TS
  23.203V16.1.0.},''

\bibitem{c14}
{\em 5GAA: Paving the Way towards 5G}, accessed on 03 September 2020.
\newblock Available online:
  \url{https://5gaa.org/5g-technology/paving-the-way}.

\end{thebibliography}

\end{document}